# Why 2 times 2 ain't necessarily 4 – at least not in IT security risk assessment

Jens Braband[1]

**Abstract:** Recently, a novel approach towards semi-quantitative IT security risk assessment has been proposed in the draft IEC 62443-3-2. This approach is analyzed from several different angles, e.g. embedding into the overall standard series, semantic and methodological aspects. As a result, several systematic flaws in the approach are exposed. As a way forward, an alternative approach is proposed which blends together semi-quantitative risk assessment as well as threat and risk analysis.

**Keywords:** IT security, risk analysis, semi-quantitative methods, risk matrix, IEC 62443.

## 1   Introduction

IEC 62443 is a new series of IT security standards for industrial control systems (ICS), which is currently being elaborated jointly by ISA and IEC. Recently, the first draft for Part 3-2, which deals with IT security risk assessment, has been circulated for comments [IEC3-2]. It contains a novel approach towards semi-quantitative IT security risk assessment, which will be discussed in this paper.

The paper is organized as follows: first, a short introduction to the general concepts of IEC 62443 is given, so that the new approach can be understood in its context. Then, this approach is analyzed formally and technically, in particular against good practices for semi-quantitative risk assessment in other sectors. By this analysis, some general flaws of the approach are revealed so that finally a discussion of how to possibly overcome these weaknesses is necessary.

## 2   Basic concepts of IEC 62443

### 2.1   Scope

A total of 12 standards or technical specifications are planned in the IEC 62443 series of standards that cover the topic of IT security for automation and control systems for industrial installations entirely and independently. This series of standards adds the topic of IT security to IEC 61508 which is the generic safety standard for programmable

---
[1] Siemens AG, Mobility Division, Ackerstrasse 22, 38126 Braunschweig, jens.braband@siemens.com

control systems. Up to now though, IEC 61508 and IEC 62443 have only been loosely linked.

IEC 62443 addresses four different aspects or levels of IT security:

- general aspects such as concepts, terminology and metrics: IEC 62443-1-x
- IT security management: IEC 62443-2-x
- system level: IEC 62443-3-x
- component level: IEC 62443-4-x

Today, however, the parts of IEC 62443 are still at different draft stages. Only a small number of parts such as IEC 62443-3-3 have already been issued as an international standard (IS) or a technical specification (TS). Due to the novelty of the IEC 62443 series in this section, the essential concepts of IEC 62443 will be explained briefly so as to improve the understanding of this paper.

**2.2    System definition**

The system and its architecture are divided into zones and conduits. The same IT security requirements apply within each zone. Every object, e.g. hardware, software or operator (e.g. administrator), shall be assigned to precisely one zone and all connections of a zone shall be identified. A zone can be defined both logically and physically. This approach matches the previous approach for railway signaling systems very well, as has been used as a basis in numerous applications [11]. Figure 1 shows a simple application of the concept, the connection of two safety zones (sharing the same security requirements) by a virtual private network (VPN) connection as the conduit.

The conduit would consist of the gateways at its borders and the connection in-between whatever the actual network would look like. Strictly speaking, management itself would be a zone with conduits connecting it with the gateways.

This example may serve as a blueprint for the connection of zones with similar IT security requirements. If zones with different IT security requirements are to be connected, different types of conduits, e.g. one-way connections or filters, have to be applied.

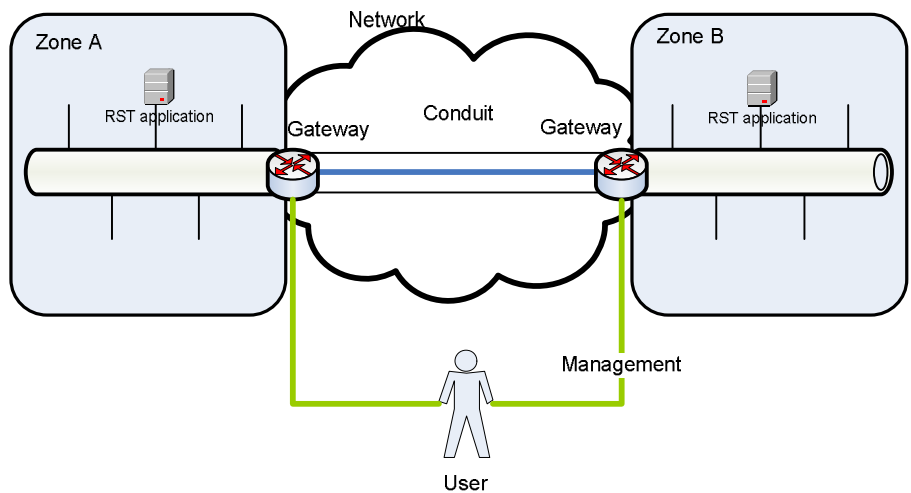

Fig. 1: Zone and conduit architecture example

## 2.3    IT security requirements

In IEC 62443, the IT security requirements are grouped into seven foundational requirements (FR):

1. identification and authentication control (IAC)
2. use control (UC)
3. system integrity (SI)
4. data confidentiality (DC)
5. restricted data flow (RDF)
6. timely response to events (TRE)
7. resource availability (RA)

Normally, only the issues of integrity, availability and data confidentiality are considered in IT security. However, the fundamental requirements IAC, UC, SI and TRE can be mapped to integrity, RA to availability, and DC and RDF to confidentiality. Instead of defining a seven-level evaluation assurance level (EAL) as in the Common Criteria, which is to be applied with regard to the IT security requirements, a four-stage IT security requirement level is defined. A possible explanation might be that also most safety standards define four levels. However, it would lead to quite demanding and sometimes unnecessary requirements if the levels were the same for each of the foundational requirements. For example, confidentiality often plays a minor role for safety systems and the encryption of all data might lead to complications in the testing or maintenance of safety systems. Hence, different levels may be assigned for each of the seven foundational requirements. The SL values for all seven basic areas are then combined into a vector, called the SL vector. Note that this theoretically leads to 16,384 possible different SLs (only partially ordered).

The SLs are defined generically in relation to the attacker type against whom they are to offer protection:

SL 1    Protection against casual or coincidental violation

SL 2    Protection against intentional violation using simple means with few resources, generic skills and a low degree of motivation

SL 3    Protection against intentional violation using sophisticated means with moderate resources, IACS-specific skills and a moderate degree of motivation

SL 4    Protection against intentional violation using sophisticated means with extended resources, IACS-specific skills and a high degree of motivation

Sometimes, a SL 0 (No protection) is also defined, but, as we argue below, at least for safety-related systems this is not an option and so we do not discuss SL 0 further in this paper.

For one zone, for example, (4, 2, 3, 1, 2, 3, 2) could be defined as an SL vector. Once its vector is defined, IEC 62443-3-3 [IEC3-3] gives a complete catalog of standardized IT security requirements for the object under consideration, e.g. for a zone.

It is necessary to take into account the fact that IEC 62443 defines different types of SL vectors:

- The target SL (SL-T) is the SL vector that results as a requirement from the IT security risk analysis.

- Achieved SL (SL-A) is the SL vector which is actually achieved in the implementation when all the particular conditions in the specific system are taken into account.

- SL capability (SL-C) is the maximum SL vector that the components or the system can reach if configured or integrated correctly, independent of the framework conditions in the specific system.

## 3  New IT security risk assessment approach

Currently, there exists no agreed approach on how to derive a SL from a threat and risk Analysis (TRA). So [IEC3-2] came up with a new approach starting with a sample risk matrix as shown in Table 1. It looks like a common approach to determine the risk R for a particular threat from the parameters likelihood L and impact I by

$$R = L \cdot I \tag{1}$$

|  | Likelihood | | | | |
|---|---|---|---|---|---|
|  | 1 Remote | 2 Unlikely | 3 Possible | 4 Likely | 5 Certain |
| **Impact** 1 Trivial | 1 | 2 | 3 | 4 | 5 |
| 2 Minor | 2 | 4 | 6 | 8 | 10 |
| 3 Moderate | 3 | 6 | 9 | 12 | 15 |
| 4 Major | 4 | 8 | 12 | 16 | 20 |
| 5 Critical | 5 | 10 | 15 | 20 | 25 |

Tab. 1: Sample risk matrix

Then, 4 is stipulated as a tolerable risk (without any justification) and any identified risk value R is then divided by 4, giving the so-called cyber security risk reduction factor (CRRF)

$$\text{CRRF} = \frac{R}{4} \tag{2}$$

and finally the target SL is derived by

$$\text{SL-T} = \min\left\{4, \left\lfloor \text{CRRF} - \frac{1}{4} \right\rfloor\right\} \tag{3}$$

A formula (3) simply states that a SL-T must not be larger than 4 and that it is more or less given by the integer part of the CRRF with a small correction of ¼. In order to understand it better, let us look at some interesting examples. For R=16, the CRRF is 4, which by (3) leads to SL-T=3. For R=17, it would lead to SL-T=4. Interestingly, both risks belong to the highest risk category in Table 1. Also, other border cases are interesting, e.g. risks labeled 6, 7 and 8 lead to SL-T=1, while 9 and 10 would result in SL-T=2. While all low-level risks should normally be acceptable, risks with 1, 2, 3, and 4 lead to SL-T=0, while 5 leads to SL-T=1. These initial observations are just an appetizer and an invitation for a more thorough analysis.

# 4     Analysis of the new approach

## 4.1     Embedding in IEC 62443

It must be clearly stated that Table 1 and (3) are only designated as examples by [IEC3-2]. But it is also clear from the process description that the example is at least meant as a blueprint. The overall process consists of the following steps

1. Identify threats
2. Identify vulnerabilities
3. Determine consequence and impact
4. Determine unmitigated likelihood
5. Calculate unmitigated cyber-security risk
6. Determine security level target
7. Identify and evaluate existing countermeasures
8. Re-evaluate likelihood and impact
9. Calculate residual risk
10. Compare residual risk with tolerable risk
11. Apply additional cyber-security countermeasures
12. Document and communicate results

As explained above, IEC 62443 derives fundamental requirements in seven different groups for zones and conduits of a particular IT security architecture, e.g. that of Figure 1. So the result should be a seven-dimensional SL-T vector instead of a scalar value given by (3). But the process description does not give any hint of how to derive the SL-T vector of a zone or conduit from the risk assessment of a threat-vulnerability combination. No explanation is given about how the concept is broken down to the foundational requirements. It may formally be argued that the authors assume that all components of the SL-T vector equal the scalar value derived by (3), but this would, in most cases, lead to very demanding requirements, e.g. for most ICS applications confidentiality is less important than integrity or availability and so the DC foundational requirement can be much weaker than that for SI or RA.

Also, at least for safety-related systems, SL-T=0 does not really make sense as protection against casual or coincidental violation should be provided in any case. It is hard to imagine a system which should not be protected against such threats. For safety-related systems, it is necessary to prevent human errors or foreseeable misuse in any case.

Additionally, there is a difference in the definition of SL between the proposal in IEC 62443-3-2 and the other parts of the standards. By applying formulae (2) and (3), the SL-T is equivalent to a risk reduction, while in the other parts, e.g. 62443-3-3, the SL-T is defined with respect to the attacker type against whom they are to offer protection. The relationship between risk reduction and the type of attacker is not explained, so it is questionable whether the approach fits to other parts of the standard.

### 4.2    Semantics

The input scales for parameters L and I are ordinal, so we know only the ordering of values 1<2<3<4<5, but have no knowledge about their further relations. For example, we do not know if an impact of 3 is five times more severe than that of 2. We could also re-label the categories to A; B, C, D, E [WP].

To make this more tangible, in programming languages such as Pascal or C, such ordinal types could be declared as

```
type
  impact = (trivial, minor, moderate, major, critical);
  likelihood = (remote, unlikely, possible, likely,
  certain);
```

Semantically, only certain operations such as predecessor, successor, ordinal number, greater than, etc., are defined for ordinal data types, but certainly not multiplication or division, which are simply undefined for ordinal data.

What is suggested by Table 1 is that the ordinal data such as "minor" is equated numerically with their order values in their type definition, e.g. `Ord(minor)` which equals 2. These order values are then treated as rational numbers and used without further explanation.

To make this argument clearer, assume that we would have labeled Table 1 with letters instead of numbers. What would $B \cdot C$ mean? Or how would the cyber-security risk reduction factor $B \cdot C/4$ be interpreted? And why should the values be multiplied and not be added?

### 4.3    Semi-quantitative risk assessment

Risk matrices like the example in Table 1 are often used in so-called semi-quantitative risk assessments such as the risk priority number (RPN) in failure modes effects and criticality assessment (FMECA). For this purpose, the classes are enriched by numerical values interpreted either as mean values for the class or as intervals.

It is well known that such approaches may have systematic flaws [Bow03], but approaches for improvement are also known [Bra03]. Here, we want to focus on the problem of multiplication only. For this discussion, we label the combination of the input parameters as the criticality of the scenario. Thus, the sample matrix in Table 1 contains the criticality numbers. The basic requirements for such approaches are:

1. If two scenarios bear a similar risk, then they should have the same criticality
2. If two scenarios are assigned to the same criticality, then they should represent similar risks

However, simple but realistic scenarios show that multiplication is not an appropriate operator with respect to these requirements. Figure 2 shows a typical example with numerical values as can also be found in standards. It can clearly be seen that both above requirements are not fulfilled as, for example, there could be scenarios that have the same criticality but the corresponding risks differ by a factor of almost one thousand. It has been shown [Bra03] that this effect is systematic and is similar for all examples. For this reason, appropriate caveats have nowadays been included in standards [IEC812].

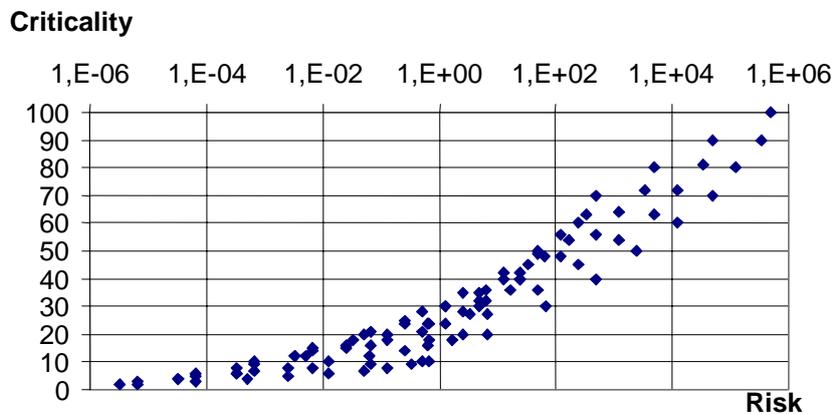

Fig. 2: Relationship between risk and criticality in RPN

So, even if the categories were defined semi-quantitatively, the basic approach as in (1) would still be flawed.

## 5   Way forward

We can summarize the analysis so far that the approach proposed by [IEC3-2] has several systematic flaws which cannot be easily overcome. In particular, the question of calculating IT security-related risks is very complex and should be avoided [BS15]. However, the use of risk matrices in IT security is so widely used in TRA that it should be kept, but it should be properly used with the definition of SL in IEC 62443.

We start from the following assumptions (without further justification):

- There exists an agreed risk matrix.
- The goal is to derive SLs which are defined by the type of attacker and the measures defined by IEC 62443.

For the sake of simplicity, we assume the same sample risk matrix as shown in Table 1 (but we do not use the criticalities). The precise form of the matrix is not important, however there should be a clear procedure which would be followed based on the color code of the results.

In a TRA, we would assess all possible threat scenarios and classify them according to their risk. The following example shows the result for three scenarios X, Y and Z. In this assessment, we have assumed that, for safety systems, we should always fulfill SL 1 =(1,1,1,1,1,1,1). This means that, in the TRA, we have assumed that all requirements related to this SL from [IEC3-3] are fulfilled, meaning that appropriate countermeasures corresponding to this SL have been implemented.

The result shows that we are OK for scenario X, but should improve for scenarios Y and Z. We now iteratively look at the assessments of the scenarios and look qualitatively for the features that should be improved from an IT security point of view. Assume in scenario Y we have a problem with authentication and user rights management. So we could increase the SL for the FR IAC and UC to (2,2,1,1,1,1,1). Let us assume this would be sufficient and we can move this scenario to the green field. But this could also have side effects on the other scenarios. Hence, we would have to re-evaluate.

|        |          | Likelihood |          |          |        |         |
|--------|----------|------------|----------|----------|--------|---------|
|        |          | Remote     | Unlikely | Possible | Likely | Certain |
| Impact | Trivial  |            |          |          |        |         |
|        | Minor    |            | X        |          |        |         |
|        | Moderate |            |          |          | Z      |         |
|        | Major    |            | Y        |          |        |         |
|        | Critical |            |          |          |        |         |

Tab. 2: Qualitative sample risk matrix

Assume in Z we have a problem with integrity, so we might also increase the SL for the FR SI to (2,2,2,1,1,1,1). If this is not sufficient, we would have to try (2,2,3,1,1,1,1) in order to make all risks acceptable after re-evaluation.

So we could use the TRA to iteratively find the SL for all zones and conduits. The TRA can be repeated at different stages in the IT security lifecycle, e.g. to determine SL-T, SL-C or SL-A.

Alternatively, we can also start to define SL by the type of attacker with another starting point, say initially SL is equal to (3,3,3,1,1,3,1), because we assumed that this level is appropriate for the foundational requirements which we have selected. Then, we would start the TRA as a check and should arrive at the same result regarding risk tolerability as before. However, as we have started with more demanding requirements, as a side effect we may also have reduced the risk associated with other scenarios, e.g. X.

## 6   Summary and conclusions

Our analysis of the approach proposed in the draft IEC 62443-3-2 has shown the following systematic flaws:

- The tolerable risk is not justified.
- The calculation of ratios used for the ordinal data is not defined.
- SL is a seven-dimensional vector, but a scalar value is derived.
- The approach contradicts the definition of SL by the type of attacker in other parts of the standard.

By way of a summary, the approach is not only unjustified, but also dangerous because the user does not need to think qualitatively about IT security; the focus is clearly on pseudo-quantification pretending unjustified accuracy.

As a way forward, an alternative approach is proposed which blends together semi-quantitative risk assessment as well as threat and risk analysis.